\documentstyle[aps,prl,twocolumn]{revtex}

\def\Journal#1#2#3#4{{#1} {\bf #2}, #3 (#4)}

\def\PRL{\em Phys. Rev. Lett.}

\def\PRB{{\em Phys. Rev.} B}
\def\SSc{\em Surf. Sci.}
\def\JPC{{\em J. Phys.} C}
\def\PR{\em Phys. Rev.}

\def\RMP{\em Rev. Mod. Phys.}
\def\PHB{\em Physica B}
\def\SSC{\em Solid State Commum.}

\input{psfig}

\begin{document}

\twocolumn[
\hsize\textwidth\columnwidth\hsize\csname@twocolumnfalse\endcsname
\draft

\title{Observation of Collective Excitations of the Dilute 2D Electron System}

\author{M.A. Eriksson$^a$, A. Pinczuk$^{a,b}$. B.S. Dennis$^a$, S.H. Simon$^a$, L.N. Pfeiffer$^a$, K.W. West$^a$}

\address{$^a$Lucent Technologies, Bell Labs, Murray Hill, NJ  07974}

\address{$^b$Departments of Physics and Applied Physics, Columbia University, New York, NY  10027}

\date{\today}

\maketitle

\begin{abstract}
  We report inelastic light scattering measurements of dispersive spin
  and charge density excitations in dilute 2D electron systems
  reaching densities less than $10^{10}~cm^{-2}$.  In the quantum Hall
  state at $\nu=2$, roton critical points in the spin inter--Landau
  level mode show a pronounced softening as $r_s$ is increased.
  Instead of a soft mode instability predicted by Hartree--Fock
  calculations for $r_s\sim 3.3$, we find evidence of multiple rotons
  in the dispersion of the softening spin excitations.  Extrapolation
  of the data indicates the possibility of an instability for $r_s
  \stackrel{>}{\scriptstyle\sim} 11$.

PACS numbers: 73.20.Mf, 73.40.Hm, 78.30.Fs
\end{abstract}
\vspace*{10pt}

]

As the density of a two-dimensional electron system decreases, unusual
behavior is expected due to the increasing importance of the
electron--electron interaction.  Wigner originally predicted that at
sufficiently low density an interaction driven phase transition to a
crystal state would occur \cite{EPW}.  It is possible, however, that
quantum phase transitions to other broken symmetry states may occur
before the onset of crystallization.  In the quantum Hall regimes, low
energy rotons in collective excitations have been identified as
possible precursors of instabilities at low electron densities
\cite{KH,MAC,GMP}.  Such collective excitations of quantum Hall
systems have been observed using inelastic light scattering
\cite{APInt,APFrac,HDM}.  Until now, however, measurements have been
restricted to high density systems, relatively far from predicted
instabilities.  In this letter we report inelastic light scattering
measurements of high mobility dilute 2D electron systems with
densities less than $10^{10}~cm^{-2}$, in the range where
instabilities have been predicted to occur \cite{MAC}.

In the quantum Hall regime breakdown of wavevector conservation due to
weak residual disorder activates light scattering by critical points
in the dispersion of collective modes, allowing the observation of
roton minima.  We find that the roton of spin--density inter--Landau
level (ILL) excitations at $\nu=2$ displays a significant softening at
low density.  At densities similar to those where Hartree-Fock
calculations predict a zero--energy spin density mode \cite{MAC},
additional peaks appear in the light scattering spectra.  These sharp
peaks can be understood as the appearance of multiple rotons in the
dispersion of the spin density excitation.  We propose that the
emergence of multiple rotons at very low density arises from a mixing
of higher Landau levels into the ground state wavefunction.  The roton
instability of Hartree-Fock calculations seems to be removed or pushed
to lower density by Landau level mixing in the ground state.

We also measure dispersive 2D plasmons (at magnetic field $B=0$) and
charge density ILL excitations at $B\neq 0$ (magnetoplasmons).  Well
defined dispersive modes are observed at densities
$n<10^{10}~cm^{-2}$.  Because of the low electron density, we are able
to measure the dispersion curves up to wavevectors $q$ approaching
$1/l_0$, where $l_{0}^{-1}=\sqrt{\hbar c/eB_\bot}$ is the magnetic
length and $B_\bot$ is the magnetic field normal to the sample.  At
low densities disorder may easily dominate the physics of the electron
system.  The observation of sharp dispersive modes is evidence that
disorder is not important in the very low density samples of this
study.

\vspace{-1.3cm}
\begin{figure}
\psfig{file=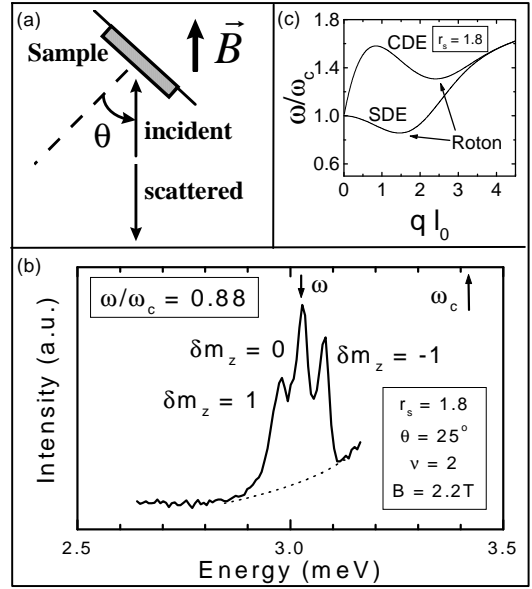,width=7cm}
\caption{(a) Experimental geometry.  Changing the angle $\theta$
  changes the scattering wavevector $q = \frac {4\pi}{\lambda}
  \sin{\theta}$.  (b) Depolarized inelastic light scattering spectrum
  of a sample with $r_s=1.8$ at $\nu=2$.  (c) Charge density and spin
  density excitations as a function of wavevector $q$ in a
  Hartree-Fock approximation. \label{fig:spinrawdata}}
\end{figure}
\vspace{-0.2cm}

The samples in this work are GaAs/Al$_x$Ga$_{1-x}$As single quantum
wells containing high quality two-dimensional electron systems.  In
order to achieve low densities without gating the samples, we use a
heterostructure design with a large setback of the quantum well from
the Si delta doping and very low aluminum concentration ($x\sim 0.03$)
in the barriers.  The transport mobility of our samples exceeds $1.5
\times 10^6$ cm$^2$/Vs at a density of $1.3 \times 10^{10}$ cm$^{-2}$.
For more dilute samples, the dispersive plasmon is used to determine
the electron density \cite{MAE}.  As shown in
Fig.~\ref{fig:spinrawdata}(a), light scattering spectra are acquired
using a backscattering geometry at sample temperatures less than 200
mK.\@ The angle $\theta$ between the normal to the sample and the
incident/scattered light is continuously tunable between $\theta = \pm
60^o$ at low temperature, allowing the scattering wavevector $q =
\frac{4\pi}{\lambda} \sin{\theta}$ to vary between $q_{min}\approx 0$
and $q_{max}=1.3 \times 10^5$ cm$^{-1}$.  Magnetic fields as high as
17 T are available, and incident power densities are always lower than
$3\times 10^{-4}$ W/cm$^2$.

It is convenient to define the electron density in terms of the usual
dimensionless parameter $r_s\equiv \frac{1}{\sqrt{\pi n}}\frac{m_b
  e^2}{\epsilon\hbar^2}$, where $m_b=0.067$ is the band effective
mass, and $\epsilon=13$ is the dielectric constant \cite{AFS}.  Large
$r_s$ corresponds to low density and strong interactions.  At $\nu=2$,
$r_s=E_c/\hbar\omega_c$, where $E_c\equiv e^2/\epsilon l_0$ is the
interaction energy, $\hbar\omega_c=\hbar eB_{\bot}/m_b$ is the
cyclotron energy, and $B_\bot = B\cos(\theta)$.

Figure \ref{fig:spinrawdata}(b) shows a light scattering spectrum of
ILL excitations in a sample with density $n = 9.6 \times 10^{10}$
cm$^{-2}$, corresponding to $r_s=1.8$.  The incident and scattered
light have perpendicular polarizations (depolarized spectra),
indicating that the light scattering peak is due to spin--density
excitations \cite{YY}.  Because of the spin symmetry at $\nu=2$, the
spin density inter--Landau level excitation is expected to be a
triplet mode \cite{KH,MAC}.  The sharp triplet of
Fig.~\ref{fig:spinrawdata}(b), whose components have a full--width at
half--maximum less than 20 $\mu eV$, is a direct observation of this
degeneracy.  The peaks in the triplet are labeled by the change in
angular momentum $\delta m_z$ parallel to the magnetic field $B$, and
their splitting is the Zeeman energy.  We observe triplet excitations
in all measured samples up to $r_s = 3.3$.  At larger $r_s$, the
Zeeman splitting is too small at $\nu=2$ to resolve the triplet.

The triplet in the depolarized spectra of
Fig.~\ref{fig:spinrawdata}(b) is shifted below the cyclotron energy
$\hbar\omega_c$ by 12\%.  The light scattering we observe in all
depolarized spectra is independent of angle $\theta$ and wavevector
$q$, so we identify the triplet as the roton minimum of the SDE, as
shown in Fig.~\ref{fig:spinrawdata}(c).  Roton minima of quantum Hall
collective modes typically occur at wavevectors larger than those
available in light scattering experiments.  Rotons have been observed,
however, because residual disorder relaxes the requirement of
wavevector conservation, allowing the observation of critical points
in the density of states \cite{APInt}.  The intensity of Raman
scattering with weak breakdown of wavevector conservation can be
written as \cite{SDS1}
\begin{equation}
I(q,\omega) \sim \int_{0}^{\infty} f_q(k) S_{a}(k,\omega )dk,
\label{breakdown}
\end{equation}
where $S_{a}(k,\omega)$ is the dynamical structure factor.  For the
charge density response $a=\rho$ and $S_\rho$ is a density--density
correlator; for the spin density response $a=\sigma$ and $S_\sigma$ is
a spin--spin correlator.  In the absence of disorder wavevector
conservation is maintained: $f_q(k)$ will be sharply peaked at $k=q$,
and $I(q,\omega)$ will map out the dispersion curves of collective
excitations.  In the presence of disorder, $f_q(k)$ is broad, and
light scattering is independent of $q$.  Then $I(\omega)$ will be
peaked at the energies of critical points in the density of states for
collective excitations, such as the roton of the SDE in
Fig.~\ref{fig:spinrawdata}(b).  In the intermediate case of very weak
disorder, both critical points in the density of states {\em and} the
wavevector conserving dispersion may be visible.

\vspace{-0.5cm}
\begin{figure}
\psfig{figure=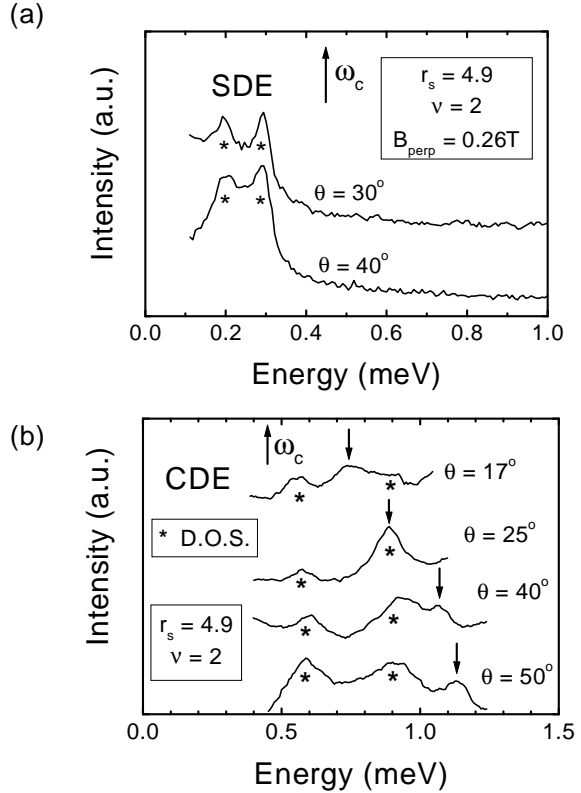,width=8.5cm}
\caption{(a)Two SDE peaks in a depolarized spectrum of a sample with $r_s=4.9$.
  The peaks are split by much more than the Zeeman energy.  (b)
  Polarized spectra from the same sample at four angles $\theta$.
  Stars indicate critical points activated by residual disorder, and
  downward arrows indicate the dispersive magnetoplasmon.
\label{fig:multipeaks}}
\end{figure}

Figure \ref{fig:multipeaks}(a) shows light scattering spectra of spin
excitations at $\nu=2$ in a sample with $r_s=4.9$.  The two sharp
peaks do not depend on angle $\theta$, indicating that they correspond
to densities of states in the dispersion curves.  Both peaks are
shifted well below $\hbar\omega_c$.  The two peaks do not correspond
to $\delta m_z$ components of the spin excitations, because the Zeeman
energy is less than $10~\mu eV$ at this field.  The appearance of two
such distinct peaks in light scattering spectra of the spin
excitations at large $r_s$ is surprising, because only a single
critical point is expected in the dispersion of spin excitations
\cite{KH,MAC,APInt}.

\begin{figure}
\psfig{figure=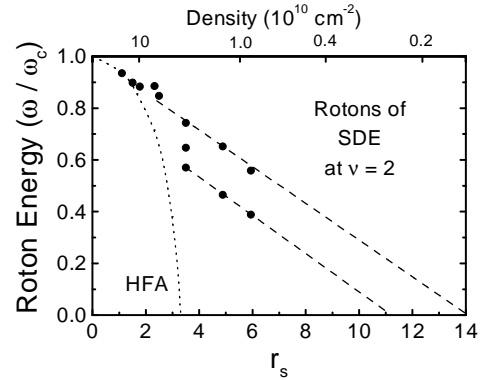,width=7cm}
\caption{{\em Solid circles:} Summary of peaks in depolarized spectra for
  eight samples, each with a different value of $r_s$.  Peaks at the
  same value of $r_s$ appear in the same sample.  Note the onset of
  multiple peaks at $r_s=3.3$ {\em Dotted line:} Calculated energy of
  the roton minimum in a Hartree-Fock approximation.  {\em Dashed
    lines:} Linear extrapolation of the data.
\label{fig:spinsummary}}
\end{figure}

The solid circles in Fig.~\ref{fig:spinsummary} are the measured
energies of spin density excitations at $\nu=2$ in 8 samples covering
a wide range in $r_s$.  We can compare these energies with a
calculation in the Hartree-Fock approximation (HFA) of the energy of
the spin--density roton minimum, shown as the dotted line in
Fig.~\ref{fig:spinsummary}.  The calculation includes both the effects
of the finite thickness of the quantum well \cite{CK} and the
self--consistent coupling between the excited states \cite{MAC}.
Although the fit is excellent for $r_s<2$, the HFA predicts a collapse
of the roton energy at $r_s\approx 3.3$ which is not observed in our
data.  The HFA is perturbatively exact when $r_s \ll 1$, but is of
questionable validity for $r_s\stackrel{>}{\scriptstyle\sim}1$,
because the Coulomb interaction becomes too strong to treat
perturbatively.

It is remarkable that the onset of multiple peaks in
spin--excitations, like the two peaks in Fig.~\ref{fig:multipeaks}(a),
occurs at approximately the same value of $r_s$ as the predicted HFA
spin instability.  Fig.~\ref{fig:spinsummary} shows that not only the
sample with $r_s=4.9$, but also the samples with $r_s=3.3$ and
$r_s=5.9$, have more than one peak in depolarized spectra.  The
multiple peaks indicate that the spin--excitation dispersion has
become more complicated at large $r_s$.  Multiple peaks can be
explained by the emergence of multiple rotons in the dispersion of the
SDE at large $r_s$, which increase the number of critical points in
the collective mode density of states.  Multiple critical points
produce multiple light scattering peaks, as can be deduced from
Eq.~(\ref{breakdown}).  A naive extrapolation of the measured roton
excitation energies to larger $r_s$, as shown by the dashed lines in
Fig.~\ref{fig:spinsummary}, reveals potential instabilities in the
range $11 \stackrel{<}{\scriptstyle\sim} r_s
\stackrel{<}{\scriptstyle\sim} 14$: although the HFA seems to
overestimate the tendency towards instability, a spin instability may
still occur before the anticipated transition to a Wigner crystal at
even larger $r_s$.

\begin{figure}
\psfig{figure=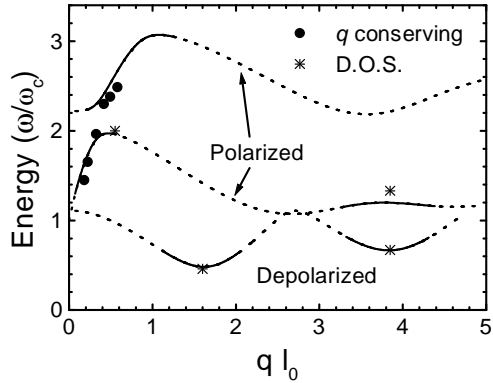,width=7.5cm}
\caption{{\em Circles}: Experimentally measured dispersion.  {\em Asterisks}:
  Energies of observed critical points.  {\em Lines:} SDE and CDE
  dispersions in a Fermi liquid theory approximation.  The solid lines
  indicate the region in the dispersions where the collective mode has
  a large weight in $S_{a}(q,\omega)$. \label{fig:128971}}
\end{figure}

Figure \ref{fig:128971} shows how the concept of multiple rotons
explains the depolarized (SDE) light scattering peaks we observe at
large $r_s$.  As shown in Fig.~\ref{fig:spinrawdata}(c), however, the
HFA predicts only a single roton minimum in the dispersion of spin
excitations.  The disagreement with Hartree-Fock calculations of the
collective modes may arise because the HFA assumes {\em a priori} that
the ground state at $\nu=2$ fills the lowest spin--split Landau level,
without Landau level mixing.  The origin of the single roton minimum
in the HFA is essentially the simple structure of the $0th$ and $1st$
(orbital) Landau level single particle wavefunctions, which have zero
and one node respectively.  We suggest that the incorporation of
higher Landau levels into the ground state at large $r_s$ increases
the complexity of the ground state wavefunction by including single
particle states with a finite number of nodes.  The finite number of
nodes can lead to multiple rotons in the collective excitations.  This
is analogous to collective excitations at higher filling factors,
which are predicted to have multiple rotons \cite{KH,MAC}.

Figure \ref{fig:multipeaks}(b) shows {\em polarized} spectra at four
angles in the sample with $r_s=4.9$.  The light scattering peaks are
due to the $\nu=2$ charge density excitation (CDE).  Two peaks, marked
by the $\star$ under the peaks, are independent of $\theta$ and are
due to critical points in the density of states (DOS) for the CDE.
Unlike all the spectra we have discussed so far, there is one peak in
Fig.~\ref{fig:multipeaks}(b) which {\it does} depend on angle $\theta$
and wavevector $q$.  It is marked by the vertical arrows in
Fig.~\ref{fig:multipeaks}(b).  At $25^o$ this peak overlaps the DOS
peak at 0.9 $meV$.  Previous experiments using inelastic light
scattering to measure the CDE in the quantum Hall regime have utilized
wavevector breakdown to measure densities of states \cite{APInt}, or
have used gratings to determine a scattering wavevector \cite{AFS,LS}.
At this density we observe both wavevector conserving {\em and}
wavevector nonconserving scattering in the same spectrum without
artificial gratings.  In polarized spectra at higher densities only
the DOS peaks are visible, and at lower densities we observe only the
dispersive mode \cite{MAE2}.

\vspace{-4.0cm}
\vspace{-0.5cm}

Figure \ref{fig:128971} is a summary of both depolarized and polarized
light scattering in the sample with $r_s=4.9$.  The four asterisks
indicate the energy of wavevector independent light scattering peaks,
which arise from densities of states in the collective modes.  The
position of the asterisks along the horizontal axis has no relation to
the scattering wavevector $q$, and serves only to elucidate the
discussion below.  The solid circles represent the wavevector
conserving peak, with $q = \frac{4\pi}{\lambda} \sin{\theta}$.

The results in Fig.~\ref{fig:128971} provide a unique opportunity to
simultaneously compare both the critical points of the dispersion
curves, and the low $q$ dispersion itself, with theoretical
calculations.  Although the HFA was used with great success to
calculate the mode dispersions at small $r_s$, it fails in the regime
of large $r_s$.  Specifically, the HFA predicts an unobserved
zero--energy roton at $r_s \approx 3.3$; it also predicts only a
single roton minimum in SDE at all values of $r_s$, whereas our
experiment shows multiple critical points.  A robust approximation for
large $r_s$ electron systems is Landau Fermi liquid theory (FLT)
\cite{PN,LQ}.  We are led to apply FLT to our experiment because it
provides a good method to treat the low density electron system at
zero or small magnetic field $B$.  In the same way that it was
instructive to push the HFA out of its range of validity to large
$r_s$, we find it useful to push Fermi liquid theory out of its strict
range of validity to large $B$, into the quantum Hall regime, in order
to compare with our results.

\vspace{-4.8cm}

The lines in Fig.~\ref{fig:128971} represent the calculated
dispersions of spin and charge modes in a semiclassical FLT
approximation using only the first significant Fermi liquid parameters
for charge, $F_{1}^{s}$, and spin, $F_{0}^{a}$ \cite{F0s}.  We recall
that $F_{1}^{s}$ and $F_{0}^{a}$ determine the effective mass and the
spin susceptibility respectively.  The best fit to our data is
obtained with $F_{1}^{s}=-0.1$ and $F_{0}^{a}=-0.87$, which are not
unreasonable values for these parameters at this density \cite{YK}.
We mark with thick lines the regions where the weight of the
collective mode is large in the dynamical structure factor
$S_{a}(q,\omega)$.  The weight in $S_{a}(q,\omega)$ oscillates because
of variations in the commensurability of the cyclotron orbits and the
wavevector of the collective oscillation; this is analogous to the
variation in the matrix elements for the overlap of particle--hole
pairs in HFA calculations.  The dispersive peak represented by the
solid circles in Fig.~\ref{fig:128971} is revealed to be the
dispersive magnetoplasmon.  The weight in $S_{a}(q,\omega)$ continues
through the avoided crossing at $\sim 2.1~meV$, as observed in the
experiment \cite{EU}.  The two critical points in the spin (charge)
mode are explained as arising from two minima (maxima) in the
dispersions.  The variation in the weight of $S_{a}(q,\omega)$
explains why peaks are observed at only specific flat regions of the
dispersion curves.

The surprisingly good numerical agreement (only two fitting parameters
were used, $F_{1}^{s}$ and $F_{0}^{a}$) between our data and
calculations based on Landau Fermi liquid theory must at this point be
regarded as fortuitous.  Neither FLT nor the HFA are fully valid in
our experimental regime.  While the HFA provides an excellent
treatment of the $\nu=2$ quantum Hall system with weak interactions,
it is expected to fail in the strong interaction limit.  Conversely,
although FLT is an excellent paradigm in which to treat strong
interactions in weak magnetic fields, it is not well suited to
handling the fully quantum limit at $\nu=2$.  These two complimentary
approaches have allowed us to suggest mixing of higher Landau levels
in the ground state wavefunction to explain our data.

In conclusion, we studied dilute 2D electron systems by means of
inelastic light scattering.  The magnetoplasmon shows a well defined
dispersion in the quantum Hall regime down to $n<10^{10}~cm^{-2}$.  At
$\nu=2$ the roton of the spin--density mode softens at large $r_s$.
The spin instability predicted by HFA calculations at $r_s\sim 3.3$ is
not observed, but we find the emergence of multiple rotons in the
spin--density mode.  We suggest a mixing of higher Landau levels in
the ground state wavefunction as an explanation for the appearance of
multiple rotons at large $r_s$.  These results are further indication
that large $r_s$ integer quantum Hall states will involve
interaction--driven correlation.  The collapsing energy of the roton
of spin excitations suggests the possibility of instabilities at large
$r_s$ due to spin correlations, in addition to the anticipated Wigner
crystal.

We wish to thank K. Todd and J.P. Eisenstein for a calculation of the
charge density in the growth direction of our quantum well samples.

\vspace{-0.5cm}


\begin{thebibliography}{99}
\bibitem{F0s} $F_{0}^{s}$ has virtually no effect on the mode
  dispersions at large $r_s$, so we neglect it in our calculations.
\end{thebibliography}
\end{document}